\documentclass[fleqn,10pt]{wlscirep}

\usepackage[utf8]{inputenc}
\usepackage[T1]{fontenc}

\usepackage{amssymb}
\usepackage{amsmath}

\usepackage{color}
\usepackage{xcolor}

\title{Metal-insulator transition of spinless fermions coupled to dispersive optical bosons}

\author[1,*]{Florian Lange}
\author[1,2,+]{Holger Fehske}
\affil[1]{Friedrich-Alexander-Universit\"at Erlangen-N\"urnberg (FAU), Erlangen National High Performance Computing Center (NHR@FAU), 91058 Erlangen, Germany}
\affil[2]{Institute of Physics, University of Greifswald, 17489 Greifswald, Germany}

\affil[*]{florian.lange@fau.de}
\affil[+]{holger.fehske@fau.de}

\begin{abstract}
Including the previously ignored dispersion of phonons we revisit the metal-insulator transition problem in one-dimensional electron-phonon systems on the basis of a modified spinless fermion Holstein model. Using matrix-product-state techniques we determine the global ground-state phase diagram in the thermodynamic limit for the half-filled band case, and show that in particular the curvature of the bare phonon band has a significant effect, not only on the transport properties characterized by the conductance and the Luttinger liquid parameter, but also on the phase space structure of the model as a whole. While a downward curved (convex) dispersion of the phonons only shifts the Tomonaga-Luttinger-liquid to charge-density-wave quantum phase transition towards stronger EP coupling, an upward curved (concave) phonon band leads to a new phase-separated state which, in the case of strong dispersion, can even completely cover the charge-density wave. Such phase separation does not occur in the related Edwards fermion-boson model.

\end{abstract}
\begin{document}

\flushbottom
\maketitle

\thispagestyle{empty}

\section*{Introduction}

Metal-insulator transitions (MITs) driven by the electron-phonon (EP) coupling have been the focus of solid-state physics studies for decades.   
The Peierls instability\cite{Pe55,Fr54} is perhaps the most prominent and fundamental example. Acting in the static (frozen-phonon)  limit, 
it establishes an insulating---charge-density-wave (CDW)---broken-symmetry state, related to a structural distortion, 
as  observed in most one-dimensional (1D) inorganic and organic conductors\cite{Gr94,Po16}. Quantum phonon fluctuations,  on the other hand, become increasingly important in low dimensions, and counteract any development of long-range order, \cite{JZW99,HF18,CBCBS18}. 

While the basic mechanisms promoting or hindering a metal-insulator quantum-phase transition are well known\cite{Mo90}, 
their detailed understanding within the framework of minimal theoretical models is challenging. 
Actually there are only a very limited number of microscopic model Hamiltonians for which such an MIT could be rigorously proven.
Holstein-like models\cite{Ho59a}. 
provide a paradigm in this respect, both in the spinless and the spinful case\cite{HF83,ZHKE23,Debnath2021}. 
By using involved numerical techniques, such as exact diagonalization and kernel polynomial\cite{WFWB00,FT07,SHBWF05,WWAF06}, diagrammatic  and quantum Monte Carlo\cite{MNP14,HFA11,WAH16}, or 
density-matrix renormalization group (DMRG)\cite{JZW99,JF07} methods, this type of model could be solved numerically 
exact in the last years. This concerns the ground-state, spectral, transport and thermodynamic properties in 1D, both for a few particles 
and for the case of the half-filled band. As a result, in the latter case, the ground-state phase diagram  of, e.g., the spinless fermion Holstein model was determined, and it has been shown that an MIT from a (repulsive) Tomonaga-Luttinger-liquid (TLL)\cite{To50,Lu63} to a CDW 
occurs when the EP coupling is increased at finite phonon frequency\cite{ZFA89,WF98b,MHM96,BMH98,HWBAF06,EF09a}. 

The original  Holstein model considers a spatially localized EP coupling to a dispersionless phonon mode\cite{Ho59a}. 
While extensions regarding the range and kind of the EP coupling and the electron hopping have been discussed  for some time\cite{AK99,FLW00,Ho16}, the influence of the missing phonon dispersion was recently questioned, but largely only for the few particle (polaron and bipolaron) problem\cite{MB13,BT21,BT22,JBH22,CF24,KB24}. Interestingly, it turned out that the phonon dispersion had a profound effect on the transport properties of Holstein and Edwards polarons in 1D. In two dimensions, it has been demonstrated  for the spinful Holstein model that the competition between pairing and charge order can be tuned by even a weak bare phonon dispersion\cite{CBCBS18}. Against this background, it seems necessary to re-examine the influence of phonon dispersion on the TLL-CDW MIT of the spinless fermion Holstein model as well. That is the main purpose of this work. To this end, we extend the model accordingly and determine the ground state and transport properties using unbiased numerical techniques for the 1D infinite half-filled Holstein system. The MIT phase boundary is compared with that of an effective electronic Hamiltonian, derived for weak phonon dispersion in Appendix A. Results for the related Edwards fermion-boson transport model~\cite{Ed06,AEF07,WFAE08,EHF09,LWF24} are presented and discussed in Appendix B.

\section*{Model and Methods}
The modified Holstein model under consideration is
\begin{align}
  \hat{H} &= -t_f \sum_j ( \hat{f}_j^{\dagger} \hat{f}_{j+1}^{\phantom{\dagger}} + \text{H.c.}) - g \omega_0 \sum_j \hat{n}_j (\hat{b}_j^\dagger + \hat{b}_j^{\phantom{\dagger}}) +  \omega_0 \sum_j \hat{b}_j^{\dagger} \hat{b}_j^{\phantom{\dagger}} + t_\omega \sum_j ( \hat{b}_j^{\dagger} \hat{b}_{j+1}^{\phantom{\dagger}} + \text{H.c.}) ,
  \label{eq:model}
\end{align}
where $\hat{f}_j$ are fermion annihilation operators describing the conduction electrons, $\hat{b}_j$ are boson annihilation operators for the phonons, and $\hat{n}_j = \hat{f}_j^\dagger \hat{f}_j^{\phantom{\dagger}}$. 
Compared with the regular Holstein model, there is an additional nearest-nearest neighbor hopping of the phonons that changes their dispersion to $\omega(k) = \omega_0 + 2 t_\omega \cos(k)$  (note the different sign convention compared with the electron hopping). To simplify the notation, we will use $t_f = 1$ as the unit of energy throughout this work. We will moreover restrict ourselves to the half-filled band case with the number of electrons equal to half the number of sites.

All our numerical simulations are based on the matrix-product-state (MPS) formalism\cite{Sch11}. 
For ground-state calculations with finite system sizes, we employ the regular DMRG algorithm\cite{Wh92}. 
Is is often more efficient, however, to work directly in the thermodynamic limit by using infinite matrix-product states (iMPS). For those simulations we apply the variational uniform MPS algorithm (VUMPS)\cite{ZVFVH18} to obtain a ground-state approximation, and the time-dependent variational principle (TDVP)\cite{HLOVV16} with bond expansion\cite{GLvD23,LGvD22} to carry out time evolutions. 
Since we consider perturbations that only affect the state in a finite region, the latter amounts to the TDVP method for finite systems with infinite boundary conditions\cite{PVM12,MHOV13,ZGEN15}. The maximum bond dimension in our simulations was $400$, except for systems with periodic boundary conditions, in which case it was $1600$. 

The infinite-dimensional local Hilbert spaces of the phonons must be truncated to a finite dimension $D_{b}$ for the numerical simulations. Depending on the model parameters, one may need a relatively large $D_b$ to accurately represent the low-energy states, particularly in the CDW phase. Several methods have been developed to make MPS simulations more efficient
for systems with large local Hilbert spaces\cite{SKMJHP21}. 
Here, we use the pseudo-site approach for the finite-system DMRG calculations\cite{JW98b}. 
For the VUMPS and TDVP simulations in the TLL phase, we did not utilize any such techniques because we found a relatively small boson dimension $D_b=8$ to be sufficient.

\section*{Results}
\subsection*{Phase diagram}
To gain insight into the general ground-state phase diagram of the model~\eqref{eq:model}, it is helpful to apply a Lang-Firsov transformation 
that changes the fermion and boson operators while leaving the fermion density operators $\hat{n}_j$ invariant\cite{LF62}. 
As shown in Appendix A, this transformation
reveals an effective electron-electron interaction that is not present in the regular Holstein model with dispersionless phonons. Since the interaction is exponentially decaying with an exponent $\text{arcosh}(\omega_0 / |2 t_\omega|)$, it suffices to consider only the nearest-neighbor part if the dispersion is small. 
Applying standard perturbation theory\cite{HF83,DDY05} then yields the following purely electronic Hamiltonian that is valid 
for $t_f/g^2, |t_\omega| \ll \omega_0$: 
\begin{align}
  \hat{H}_{\text{eff}} &= -\tilde{t}_f  \sum_j \left( \hat{f}_j^{\dagger} \hat{f}_{j+1}^{\phantom{\dagger}} + \text{H.c.} \right) + \Big[\frac{\tilde{t}_f^2} {\omega_0}  \big(4 g_1 + 2 g_2 \big)  + 2 t_\omega g^2 \Big] \sum_j  \hat{n}_j \hat{n}_{j+1}
  + \frac{\tilde{t}_f^2 }{\omega_0}  g_1 \sum_j \big[ \hat{f}_{j-1}^{\dagger} (2 \hat{n}_j -1) \hat{f}_{j+1}^{\phantom{\dagger}} + \text{H.c.} \big] \,,
  \label{eq:Heff}
\end{align}
where $\tilde{t}_f=t_f e^{-a^2}$, $g_1 = \sum_{n=1}^\infty \frac{a^{2n}}{n!n}$, $g_2 = \sum_{n=1}^\infty \sum_{m=1}^\infty \frac{a^{2(n+m)}}{n!m!(n+m)}$, and
$a = g(1 - t_\omega / \omega_0)$. 
The main difference compared with the effective model for the regular Holstein model is the additional contribution $\propto t_\omega $ to the interaction term. From Eq.~\eqref{eq:Heff} one can draw some qualitative conclusions about the phase diagram. For $t_\omega = 0$, the interaction is always repulsive, and since it decreases slower with $g$ than the effective hopping strength, it causes a transition from a TLL to a CDW as $g$ is increased\cite{HF83}. When $t_\omega$ is finite, however, there is an additional term that grows quadratically with $g$ and therefore becomes decisive at strong EP coupling. In particular, it leads to an attractive interaction if $t_\omega < 0$, which indicates the possibility of an attractive TLL, and even phase separation, both of which do not occur in the regular Holstein model. For $t_\omega > 0$, on the other hand, the phonon dispersion should simply move the CDW-TLL transition to lower EP couplings. 
Since it is caused by a strong nearest-neighbor attraction, the phase-separated state considered here consists of two subsystems that are empty and occupied, respectively, with small particle-number fluctations at the interface. For open boundary conditions, the electrons will accumulate on either the left or the right side of the system, so that the number of bonds where both sites are empty or both sites are occupied is maximized.

\begin{figure}[bt]
\centering
\includegraphics[width=\linewidth]{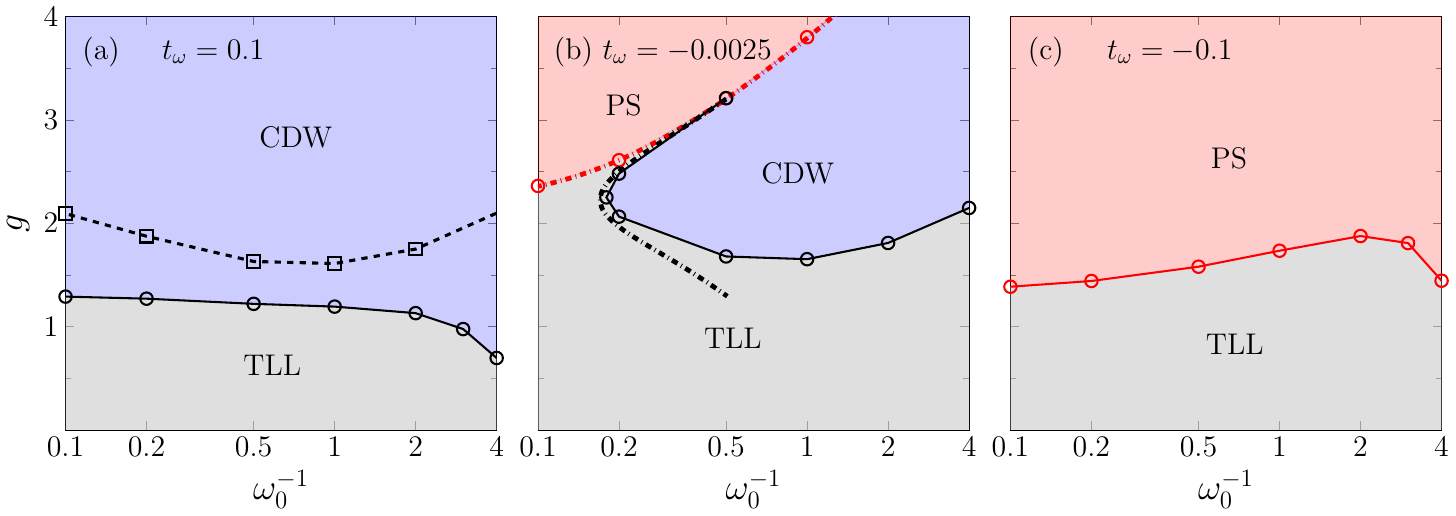}
	\caption{Ground-state phase diagram for different values of the phonon hopping $t_\omega$. Circles mark the numerically calculated phase boundaries, with the solid lines as a guide to the eye. 
	For comparison, the dashed line in (\textbf{a}) shows the TLL-CDW boundary in the dispersionless Holstein model\cite{EF09a}. 	
	The dash-dotted lines in (\textbf{b}) indicate the transitions according to the effective model~\eqref{eq:Heff} when neglecting the correlated next-nearest-neighbor hopping. }
\label{fig:pd}
\end{figure}

Since the effective Hamiltonian~\eqref{eq:Heff} has a limited region of validity, it is important to check the above predictions with numerical calculations. 
Let us first discuss the resulting phase diagram shown in Fig.~\ref{fig:pd}. 
For $t_\omega = 0.1$, there is indeed a significant shift of the TLL-CDW transition to lower values of $g$ for all $\omega_0$. 
The stronger effect in the adiabatic regime is likely related to the fact that we consider a fixed $t_\omega$, so that the effective interaction becomes longer ranged at small $\omega_0$. 
Going to a small negative $t_\omega = -0.0025$, a region with phase separation appears at large $g$. It is mostly located above the CDW phase, an exception being the anti-adiabatic regime, where the CDW phase disappears and there is a direct transition from a TLL to phase separation. 
The phase-transition points at strong-coupling can also be estimated by dropping the last term in Eq.~\eqref{eq:Heff}, which leaves only a $t$-$V$ Hamiltonian whose phase diagram is known exactly\cite{Gi03}. As demonstrated in Fig.~\ref{fig:pd}(\textbf{b}), the phase-separation boundary obtained this way agrees quite well with that of the full model. 
Interestingly, the competition between the effective repulsive interaction from perturbation theory and the attractive interaction due to the phonon dispersion causes a reentrant CDW-TLL transition that is visible at $\omega_0 = 5$. When $\omega_0 $ is lowered, this second TLL region becomes smaller and effectively disappears. 
Already at moderate negative hopping $t_\omega = -0.1$, we no longer observe a CDW phase, i.e., the system appears to always be in an attractive TLL or phase-separated state. We would like to point out that, in contrast, the half-filled Edwards model with dispersive bosons neither forms an attractive TLL nor a phase-separated state  for negative values of $t_\omega$, see Appendix B.

\begin{figure}[tb]
\centering
\includegraphics[width=\linewidth]{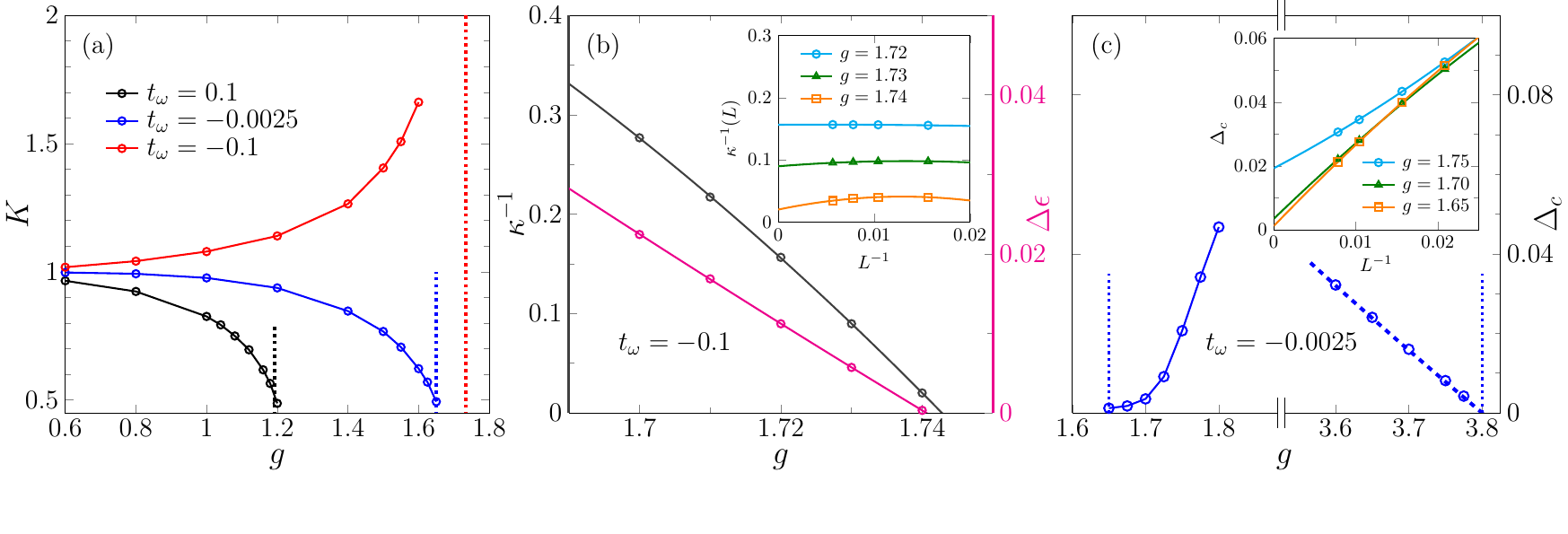}
\caption{(\textbf{a}) TLL parameter $K$ as a function of the EP coupling $g$ for $\omega_0=1$ and different phonon dispersions. Dotted lines mark the locations of phase transitions. (\textbf{b}) Inverse compressibility $\kappa^{-1}$ and ground-state energy per site $\Delta \epsilon$ relative to a phase-separated state for $\omega_0=1$ and $t_\omega=-0.1$. The inset shows the extrapolation from finite-system DMRG calculations of $\kappa^{-1}(L)$ with open boundary conditions. (\textbf{c}) Charge gap in the CDW phase for $\omega_0=1$ and $t_\omega = -0.1$. The dashed line shows the charge gap when only the nearest-neighbor-repulsion term is considered in Eq.~\eqref{eq:Heff}. The calculations were done for finite systems with open boundary conditions and then extrapolated to the thermodynamic limit using a second-order polynomial. For parameters near the TLL-CDW transition the finite-size extrapolation is presented in the inset.  }
\label{fig:K}
\end{figure}

Within the TLL phase, the system is characterized by the TLL parameter $K$, which can be used to distinguish between effective repulsive $(K < 1)$ and attractive interactions $(K > 1)$. 
Furthermore, $K=1/2$ signals a transition to the CDW phase, while $K \to \infty$ indicates the onset of phase separation. 
There are several ways to numerically determine the TLL parameter $K$. 
Here, we use its relation to the linear conductance\cite{KFM92,KLOKC21} and the charge structure factor\cite{GS89,EGN05,KM12}. 
Figure~\ref{fig:K} displays the results for an intermediate phonon energy $\omega_0=1$. 
At weak EP coupling, the TLL parameter is close to the value $K=1$ of a non-interacting chain of fermions regardless of the phonon parameters. 
As the EP coupling $g$ increases, however, the effect of the phonon dispersion on $K$ becomes significant. 
For $t_\omega=0.1$ and $t_\omega=-0.0025$, $K$ decreases with $g$ up to the TLL-CDW transition where $K=1/2$, indicating that the TLL for these parameters is repulsive as in the regular Holstein model. In contrast, setting $t_\omega = -0.1$ clearly leads to an attractive TLL with $K > 1$, similar to a sufficiently strong longer-ranged EP coupling\cite{HAF12}.  
The rapid increase of $K$ near $g \approx 1.5$ also suggests that a phase-separation transition is approached. However, while calculating $K$ allows to accurately locate the TLL-CDW transition, it is difficult to prove the occurrence of phase separation in this manner. 
We therefore also compute the inverse compressibility, 
\begin{align}
  \kappa^{-1}(L) &= \frac{L}{2}\left[E_0(L/2+1) + E_0(L/2-1) - 2 E_0(L/2)\right],
  \label{kappainv}
\end{align}
where $E_0(N)$ is the ground-state energy for a system with $N$ electrons, and $L$ is the number of sites\cite{OLSA91,ESBF12}. 
For $\omega_0 = 1$ and $t_\omega = -0.1$, $\kappa^{-1} = \lim_{L \rightarrow \infty}\kappa^{-1}(L)$ vanishes around $g \approx 1.74$, which indicates that the system indeed becomes unstable towards phase separation at this point. Furthermore, we found no evidence for electron pairing, i.e., defining $\kappa^{-1}(L)$ in terms of the two-particle gap instead of the single-particle gap as in Eq.~\eqref{kappainv} does not lead to a smaller $\kappa^{-1}$ for the parameters considered here. 
As an alternative to the inverse compressibility, one can also look at the ground-state energy per site $\epsilon_0$ from an iMPS simulation and extrapolate where it becomes larger than that of a phase-separated state with an empty and a fully occupied subsystem, i.e., where $\Delta \epsilon =  - \frac{1}{2} g^2 \omega_0^2 / (\omega_0 + 2 t_\omega) - \epsilon_{0}  = 0$. 
The transition point obtained this way agrees reasonably well with that from the inverse compressibility, which shows that phase separation indeed occurs between an empty and an occupied phase. 
We therefore use this simpler method to determine the remaining phase-separation boundaries.

The TLL-CDW transition can also be determined by locating the breakdown of the CDW state, which is
signaled by a closing of the charge gap $\Delta_c = E_0(L/2+1) + E_0(L/2-1) - 2 E_0(L/2)$. 
Figure~\ref{fig:K}(\textbf{c}) shows $\Delta_c$ in the CDW phase near the phase transitions for model parameters $\omega_0 = 1$ and $t_\omega=-0.0025$. 
For EP couplings $g$ close to the lower phase boundary, the results are clearly consistent with those for the TLL parameter $K$, 
although accurately determining the transition point this way is difficult because of the exponential closing of the gap at the Kosterlitz-Thouless transition. 
Near the second transition, on the other hand, the electron motion is almost completely frozen because of the exponential decrease of the effective hopping amplitude $\tilde{t}_f$ with the EP coupling $g$, so that the charge gap in this region is approximately proportional to the coefficient of the nearest-neighbor interaction in Eq.~\eqref{eq:Heff}. 
The closing of the gap coincides with $\Delta \epsilon$ becoming zero, which suggests that only a vanishingly small intermediate TLL phase exists.

\begin{figure}[tb]
\centering
\includegraphics[width=0.9\linewidth]{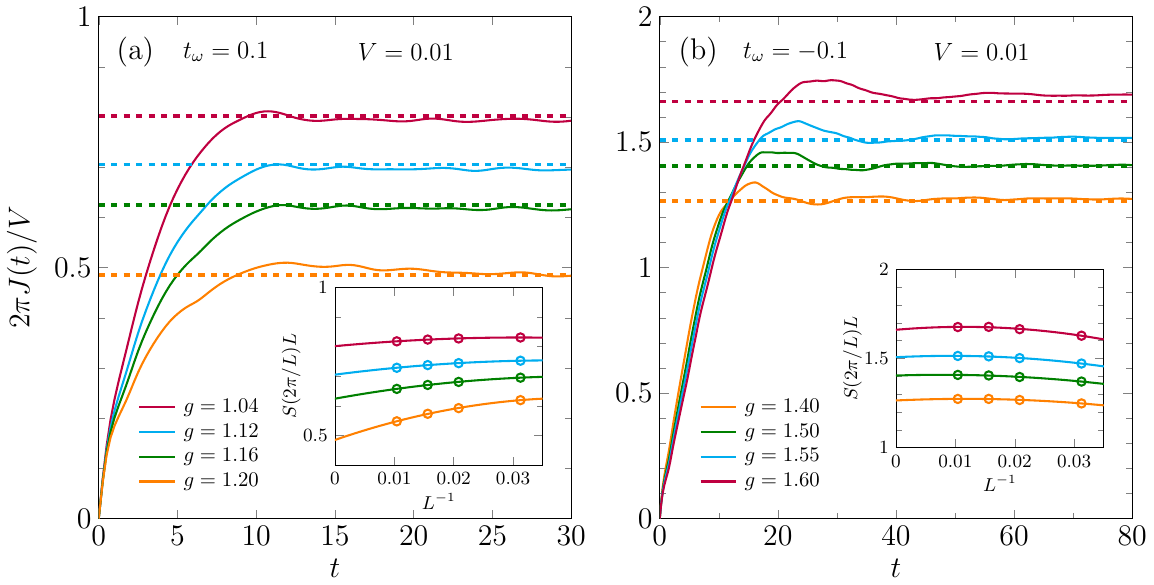}
\caption{(\textbf{a}), (\textbf{b}) Examples for the calculation of $K$ by means of the charge conductance for $\omega_0=1$. The inset of shows an alternative method based on the charge structure factor in finite systems with periodic boundary conditions. In the main panels, the extrapolated values of $K$ are displayed as dashed lines.} 
\label{fig:J}
\end{figure}

\subsection*{Calculation of the Tomonaga-Luttinger parameter $K$}
In the TLL phase the linear conductance at zero temperature is given by $K / (2 \pi)$\cite{KFM92}. Accordingly, one can determine the TLL parameter from the charge-current response to a small voltage $V$ as 
\begin{align}
K &= \lim_{t \rightarrow \infty} \lim_{V \rightarrow 0} \frac{2 \pi J(t)}{V} .
\end{align}
The charge current at time $t$ for an arbitrary bond $j$ is $J_j(t) = -it_f \langle \hat{f}_j^\dagger \hat{f}_{j+1}^{\phantom{\dagger}} - \hat{f}_{j+1}^\dagger \hat{f}_j^{\phantom{\dagger}} \rangle_t $. 
Specifically, we apply a constant potential gradient over $L_V = 10$ sites and calculate the average current $J(t)$ in that region.
The corresponding perturbation to the Hamiltonian for times $t>0$ is $\sum_j V_j \hat{n}_j$, with $V_j = \min(\max(-\tfrac{V}{2} + \tfrac{jV}{L_V+1},-\tfrac{V}{2}),\tfrac{V}{2})$. 
 Figure~\ref{fig:J} shows the obtained current $J(t)$ for parameters near the phase transitions for $\omega_0 = 1$ and $t_\omega = \pm 0.1$.
 Except for small fluctuations, $J(t)$ saturates with time, which allows to estimate the conductance and thereby $K$. Only for $t_\omega = 0.1$ and $g = 1.2$, the current $J(t)$ appears to decrease at long times, likely because the system is already slightly in the insulating CDW regime.

Another way to determine the TLL parameter $K$ is the relation\cite{EGN05}
\begin{align}
  K &= \lim_{q \rightarrow 0 } \frac{2 \pi S(q)}{q} \,,
\end{align}
which connects $K$ to the static structure factor $S(q)= \frac{1}{L}\sum_{j \ell} e^{i q(j - \ell)} \langle \hat{n}_j \hat{n}_\ell \rangle$. 
Although $S(q)$ can be calculated from the iMPS approximation of the ground state, we found that using finite systems with periodic boundary conditions and $K = \lim_{L \rightarrow \infty} L S(2 \pi /L)$ leads to a weaker momentum dependence and thus a clearer extrapolation. As demonstrated in Fig.~\ref{fig:J}, the TLL parameters obtained this way are consistent with the charge current at long times. We also used the approach based on the structure factor for the $3$ points in Fig.~\ref{fig:pd}(\textbf{b}) around $\omega_0=5$, where we encountered convergence issues in the iMPS method. 

\newpage

\section*{Conclusions}
We used MPS techniques to numerically investigate the effect of a finite phonon dispersion on the phase diagram of the one-dimensional Holstein model, focusing in particular on the TLL parameter $K$, which we have extracted from both static and dynamic quantities. 
In agreement with the derived effective strong-coupling Hamiltonian, our results demonstrate that a downward dispersion increases the tendency to CDW order, while
an upward dispersion leads to an effective electron-electron attraction. 
As the EP coupling is increased, the attractive interaction manifests itself first as a TLL with $K > 1$, and then as phase separation between an empty and a full subsystem. 
This is markedly different from the phase diagram of the Holstein model without phonon dispersion that consists only of a repulsive TLL and a CDW, but resembles the situation in models with longer-ranged electron-phonon coupling\cite{HAF12}. As shown in Appendix B, the corresponding Edwards fermion-boson model also shows no attractive TLL or phase separation.

While we studied the half-filled case in this work, it would also be interesting to investigate the phase diagram at other densities. For example, a downward phonon dispersion and strong EP coupling should lead to phase separation between a CDW and an empty region, since the effective interaction in that case favours a CDW with wave number $\pi$.
Lastly, it might be worthwhile to consider extensions of the model with other dispersion or more general types of EP coupling. A natural question in this regard is, whether the different effective electron-electron interactions can lead to more exotic phases, such as TLLs with electron pairing, which were observed in purely fermionic chains with specific longer-ranged interactions %\cite{GMSR21a}
or pair-hopping terms\cite{GMSR21b}.

\section*{Appendix A: effective electronic Hamiltonian}
In the Holstein model with boson dispersion $\omega(k)$, the local electron-phonon coupling term can be removed by a Lang-Firsov transformation $\hat{H} \to e^{\hat{S}} \hat{H} e^{-\hat{S}}$ with $\hat{S} = -\tfrac{g}{\sqrt{N}} \sum_j \sum_k \omega(k)^{-1} \big[ e^{-ikj} \tilde{b}_k^\dagger - e^{ikj} \tilde{b}_k^{\phantom{\dagger}} \big] \hat{n}_j$\cite{LF62}.
Here, we used the boson operators $\tilde{b}_k = \tfrac{1}{\sqrt{N}} \sum_j e^{-ikj} \hat{b}_j$ in momentum space. 
The transformed Hamiltonian is
\begin{align}
  \hat{H} &=  -t_f e^{-g^2 \sum_{r>0} (F_r - F_{r-1})^2} \sum_j \left[ \hat{f}_j^{\dagger} \hat{f}_{j+1}^{\phantom{\dagger}}  e^{-g \sum_{r\in \mathbb{Z}} F_r \left(\hat{b}_{j+r}^\dagger - \hat{b}_{j+1+r}^\dagger \right) } e^{g \sum_{r\in \mathbb{Z}} F_r\left( \hat{b}_{j+r}^{\phantom{\dagger}} -   \hat{b}_{j+1+r}^{\phantom{\dagger}} \right) } + \text{H.c.} \right] \nonumber \\ & \hspace*{0.4cm} +  \sum_j \sum_{r = 1}^\infty V_{r}  \hat{n}_j \hat{n}_{j+r} + \sum_k \omega(k) \tilde{b}_k^{\dagger} \tilde{b}_k^{\phantom{\dagger}} ,
\label{eq:HLangFirsov}
\end{align}
where $F_r = \frac{\omega_0}{2\pi}\int_{-\pi}^{\pi}  \frac{\cos(k r)}{\omega_0 + 2 t_\omega \cos(k)} dk = \frac{1}{\text{tanh}(\alpha)} \big[{-}\text{sgn}(t_\omega / \omega_0)\big]^r  e^{-\alpha |r|}$ with $\alpha = \text{arcosh}(\omega / |2 t_\omega|)$.
The most significant effect of the boson dispersion is the electron-electron interaction with coefficients $V_r = 2 \omega_0 g^2  F_r$. For small dispersion $|t_\omega| \ll \omega_0$, only the nearest-neighbor term with $V_1 \approx 2 t_\omega g^2$ is important, which is repulsive (attractive) for positive (negative) $t_\omega$. As in the model without dispersion\cite{HF83,DDY05}, one can apply second-order perturbation theory to Eq.\eqref{eq:HLangFirsov} in order to obtain an effective strong-coupling Hamiltonian for $t_f \ll \omega_0 g^2$. 
The calculation is complicated by the fact that the hopping term for a given electron site involves boson operators over a larger spatial range than in the model without dispersion. For $|t_\omega| \ll \omega_0$, however, we can make the approximations
$e^{-g \sum_{r\in \mathbb{Z}} F_r (\hat{b}_{j+r}^\dagger - \hat{b}_{j+1+r}^\dagger ) } e^{g \sum_{r\in \mathbb{Z}} F_r ( \hat{b}_{j+r}^{\phantom{\dagger}} -   \hat{b}_{j+1+r}^{\phantom{\dagger}} ) } \approx  e^{-g(F_0 - F_1) ( \hat{b}_{j}^\dagger - \hat{b}_{j+1}^\dagger ) } e^{g (F_0 - F_1) ( \hat{b}_{j}^{\phantom{\dagger}} -   \hat{b}_{j+1}^{\phantom{\dagger}} ) }$ and $F_1 - F_0 \approx 1 - t_\omega / \omega_0$, which results in Eq.~\eqref{eq:Heff} of the main text. 
The effective Hamiltonian~\eqref{eq:Heff} includes a repulsive electron-electron interaction term that corresponds to a second-order process where an electron hops back and forth between two sites. Unlike the terms with coefficients $V_r$, this interaction is also present for $t_\omega=0$.

\section*{Appendix B: Edwards fermion-boson model}
\begin{figure}[h]
\centering
\includegraphics[width=0.9\linewidth]{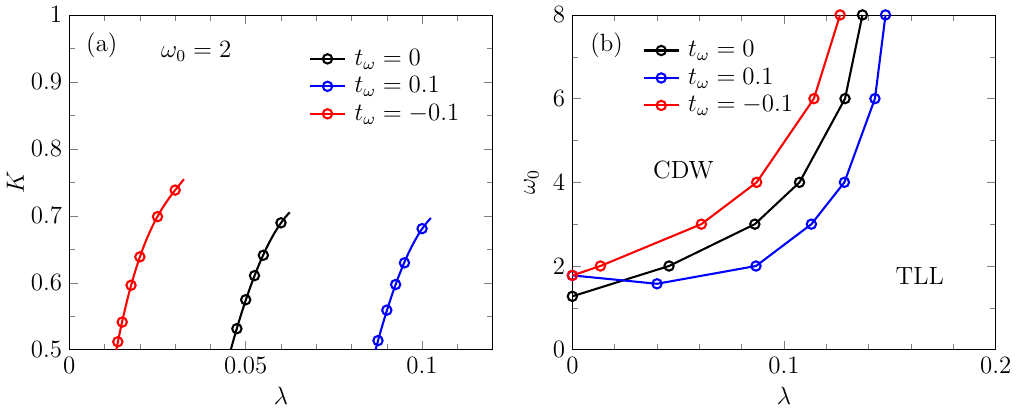}
\caption{(\textbf{a}) TLL parameter in the Edwards model with dispersive bosons and (\textbf{b}) the corresponding ground-state phase diagram.} 
\label{fig:edwards}
\end{figure}

The Hamiltonian of the 1D modified Edwards model,
\begin{align}
  \hat{H} &= -t_{fb} \sum_{j} \left( \hat{f}_{j+1}^\dagger \hat{f}_{j}^{\phantom{\dagger}} ( \hat{b}_{j}^\dagger + \hat{b}_{j+1}^{\phantom{\dagger}}) + \text{H.c.} \right) 
   - \lambda \sum_j (\hat{b}_j^\dagger + \hat{b}_j^{\phantom{\dagger}}) + \omega_0 \sum_j \hat{b}_j^\dagger \hat{b}_j^{\phantom{\dagger}} +   t_\omega \sum_j ( \hat{b}_j^{\dagger} \hat{b}_{j+1}^{\phantom{\dagger}} + \text{H.c.})  ,
  \label{eqHam}
\end{align}
describes quantum transport in a background medium parametrized by dispersive bosons.
Here, a fermion $f$ emits or absorbs a boson $b$ of energy $\omega(k)=\omega_0+2t_\omega\cos(k)$ every time it hops between neighboring lattice sites\cite{Ed06,CF24}. The bosons can also relax via the $\lambda$-term without fermion hopping. 
When the bosons are dispersionless, a TLL-CDW transition was found for a half-filled band in one dimension~\cite{WFAE08,EHF09,LWF24}. 
In the Edwards model, the insulating phase is realized for a stiff background, i.e., at small $\lambda$ and large $\omega_0$. 
  The question now is how the dispersion of the bosons influences the position of the MIT phase boundary and in particular whether an attractive TLL or phase separation also occurs in the Edwards model when $t_\omega<0$. 
To address this, we use the same numerical approach as for the Holstein model and determine the TLL parameter $K$ by calculating the charge current due to a small voltage $V=0.01$, using $t_{fb}$ as the unit of energy.
  Figure~\ref{fig:edwards}(\textbf{a}) displays the obtained $K$ values for $\omega_0 = 2$ and $t_\omega = \pm 0.1$.
  The TLL-CDW transition, where $K=1/2$, is significantly shifted by the added boson dispersion. Obviously,the CDW region gets--by and large--bigger for positive $t_\omega$ (cf. data for $t_\omega=0.1$) 
  while it becomes smaller for negative $t_\omega$ (see results for -0.1). 
  As shown in Fig.~\ref{fig:edwards}(\textbf{b}), this trend also applies to other boson energies $\omega_0$. The only exception is the region of small $\omega_0$ and $\lambda$, where both the downward and upward boson dispersions reduce the CDW region. In fact, the sign of the boson hopping does not affect the phase at $\lambda = 0$, since there it can be changed by a gauge transformation of the fermions and bosons $(\hat{f}_j,\hat{b}_j) \rightarrow (i^{j^2} \hat{f}_j,i(-1)^{j+1}\hat{b}_j)$. The most important difference to the spinless fermion Holstein model, however, is the absence of phase separation for $t_\omega = -0.1$.

\section*{Data availability}
The data that support the findings of this study are available from the corresponding author upon reasonable request.

%\bibliography{ref_16_1_24_cf24}

\begin{thebibliography}{10}
\urlstyle{rm}
\expandafter\ifx\csname url\endcsname\relax
  \def\url#1{\texttt{#1}}\fi
\expandafter\ifx\csname urlprefix\endcsname\relax\def\urlprefix{URL }\fi
\expandafter\ifx\csname doiprefix\endcsname\relax\def\doiprefix{DOI: }\fi
\providecommand{\bibinfo}[2]{#2}
\providecommand{\eprint}[2][]{\url{#2}}

\bibitem{Pe55}
\bibinfo{author}{Peierls, R.}
\newblock \emph{\bibinfo{title}{Quantum Theory of Solids}}
  (\bibinfo{publisher}{Oxford University Press}, \bibinfo{address}{Oxford},
  \bibinfo{year}{1955}).

\bibitem{Fr54}
\bibinfo{author}{Fr{\"o}hlich, H.}
\newblock \bibinfo{journal}{\bibinfo{title}{Electrons in lattice fields}}.
\newblock {\emph{\JournalTitle{Adv. Phys.}}} \textbf{\bibinfo{volume}{3}},
  \bibinfo{pages}{325} (\bibinfo{year}{1954}).

\bibitem{Gr94}
\bibinfo{author}{Gr{\"u}ner, G.}
\newblock \emph{\bibinfo{title}{Density Waves in Solids}}
  (\bibinfo{publisher}{Addison Wesley, Reading, MA}, \bibinfo{year}{1994}).

\bibitem{Po16}
\bibinfo{author}{Pouget, J.-P.}
\newblock \bibinfo{journal}{\bibinfo{title}{The {P}eierls instability and
  charge density wave in one-dimensional electronic conductors}}.
\newblock {\emph{\JournalTitle{C. R. Phys.}}} \textbf{\bibinfo{volume}{17}},
  \bibinfo{pages}{332} (\bibinfo{year}{2016}).

\bibitem{JZW99}
\bibinfo{author}{Jeckelmann, E.}, \bibinfo{author}{Zhang, C.} \&
  \bibinfo{author}{White, S.~R.}
\newblock \bibinfo{journal}{\bibinfo{title}{Metal-insulator transition in the
  one-dimensional {Holstein} model at half filling}}.
\newblock {\emph{\JournalTitle{Phys. Rev. B}}} \textbf{\bibinfo{volume}{60}},
  \bibinfo{pages}{7950} (\bibinfo{year}{1999}).

\bibitem{HF18}
\bibinfo{author}{Hohenadler, M.} \& \bibinfo{author}{Fehske, H.}
\newblock \bibinfo{journal}{\bibinfo{title}{Density waves in strongly
  correlated quantum chains}}.
\newblock {\emph{\JournalTitle{Eur. Phys. J. B}}}
  \textbf{\bibinfo{volume}{91}}, \bibinfo{pages}{204} (\bibinfo{year}{2018}).

\bibitem{CBCBS18}
\bibinfo{author}{Costa, N.~C.}, \bibinfo{author}{Blommel, T.},
  \bibinfo{author}{Chiu, W.-T.}, \bibinfo{author}{Batrouni, G.} \&
  \bibinfo{author}{Scalettar, R.~T.}
\newblock \bibinfo{journal}{\bibinfo{title}{Phonon dispersion and the
  competition between pairing and charge order}}.
\newblock {\emph{\JournalTitle{Phys. Rev. Lett.}}}
  \textbf{\bibinfo{volume}{120}}, \bibinfo{pages}{187003}
  (\bibinfo{year}{2018}).

\bibitem{Mo90}
\bibinfo{author}{Mott, N.~F.}
\newblock \emph{\bibinfo{title}{Metal-Insulator Transitions}}
  (\bibinfo{publisher}{Taylor \& Francis}, \bibinfo{address}{London},
  \bibinfo{year}{1990}).

\bibitem{Ho59a}
\bibinfo{author}{Holstein, T.}
\newblock \bibinfo{journal}{\bibinfo{title}{Studies of polaron motion. {Part
  I}. {The} molecular-crystal model}}.
\newblock {\emph{\JournalTitle{Ann. Phys. (N.Y.)}}}
  \textbf{\bibinfo{volume}{8}}, \bibinfo{pages}{325} (\bibinfo{year}{1959}).

\bibitem{HF83}
\bibinfo{author}{Hirsch, J.~E.} \& \bibinfo{author}{Fradkin, E.}
\newblock \bibinfo{journal}{\bibinfo{title}{Phase diagram of one-dimensional
  electron-phonon systems. {II}. {T}he molecular-crystal model}}.
\newblock {\emph{\JournalTitle{Phys. Rev. B}}} \textbf{\bibinfo{volume}{27}},
  \bibinfo{pages}{4302} (\bibinfo{year}{1983}).

\bibitem{ZHKE23}
\bibinfo{author}{Zhao, S.}, \bibinfo{author}{Han, Z.},
  \bibinfo{author}{Kivelson, S.~A.} \& \bibinfo{author}{Esterlis, I.}
\newblock \bibinfo{journal}{\bibinfo{title}{One-dimensional {H}olstein model
  revisited}}.
\newblock {\emph{\JournalTitle{Phys. Rev. B}}} \textbf{\bibinfo{volume}{107}},
  \bibinfo{pages}{075142} (\bibinfo{year}{2023}).

\bibitem{Debnath2021}
\bibinfo{author}{Debnath, D.}, \bibinfo{author}{Malik, M.~Z.} \&
  \bibinfo{author}{Chatterjee, A.}
\newblock \bibinfo{journal}{\bibinfo{title}{A semi exact solution for a
  metallic phase in a {H}olstein-{H}ubbard chain at half filling with
  {G}aussian anharmonic phonons}}.
\newblock {\emph{\JournalTitle{Scientific Reports}}}
  \textbf{\bibinfo{volume}{11}}, \bibinfo{pages}{12305} (\bibinfo{year}{2021}).

\bibitem{WFWB00}
\bibinfo{author}{Wei{\ss}e, A.}, \bibinfo{author}{Fehske, H.},
  \bibinfo{author}{Wellein, G.} \& \bibinfo{author}{Bishop, A.~R.}
\newblock \bibinfo{journal}{\bibinfo{title}{Optimized phonon approach for the
  diagonalization of electron-phonon problems}}.
\newblock {\emph{\JournalTitle{Phys. Rev. B}}} \textbf{\bibinfo{volume}{62}},
  \bibinfo{pages}{R747} (\bibinfo{year}{2000}).

\bibitem{FT07}
\bibinfo{author}{Fehske, H.} \& \bibinfo{author}{Trugman, S.~A.}
\newblock \bibinfo{title}{Numerical solution of the {H}olstein polaron
  problem}.
\newblock In \bibinfo{editor}{Alexandrov, A.~S.} (ed.)
  \emph{\bibinfo{booktitle}{Polarons in Advanced Materials}}, vol.
  \bibinfo{volume}{103} of \emph{\bibinfo{series}{Springer Series in Material
  Sciences}}, \bibinfo{pages}{393--461} (\bibinfo{publisher}{Canopus/Springer
  Publishing}, \bibinfo{address}{Dordrecht}, \bibinfo{year}{2007}).

\bibitem{SHBWF05}
\bibinfo{author}{Sykora, S.}, \bibinfo{author}{H{\"u}bsch, A.},
  \bibinfo{author}{Becker, K.~W.}, \bibinfo{author}{Wellein, G.} \&
  \bibinfo{author}{Fehske, H.}
\newblock \bibinfo{journal}{\bibinfo{title}{Single-particle excitations and
  phonon softening in the one-dimensional spinless {H}olstein model}}.
\newblock {\emph{\JournalTitle{Phys. Rev. B}}} \textbf{\bibinfo{volume}{71}},
  \bibinfo{pages}{045112} (\bibinfo{year}{2005}).

\bibitem{WWAF06}
\bibinfo{author}{Wei{\ss}e, A.}, \bibinfo{author}{Wellein, G.},
  \bibinfo{author}{Alvermann, A.} \& \bibinfo{author}{Fehske, H.}
\newblock \bibinfo{journal}{\bibinfo{title}{The kernel polynomial method}}.
\newblock {\emph{\JournalTitle{Rev. Mod. Phys.}}}
  \textbf{\bibinfo{volume}{78}}, \bibinfo{pages}{275} (\bibinfo{year}{2006}).

\bibitem{MNP14}
\bibinfo{author}{Mishchenko, A.~S.}, \bibinfo{author}{Nagaosa, N.} \&
  \bibinfo{author}{Prokof'ev, N.}
\newblock \bibinfo{journal}{\bibinfo{title}{Diagrammatic {M}onte {C}arlo method
  for many-polaron problems}}.
\newblock {\emph{\JournalTitle{Phys. Rev. Lett.}}}
  \textbf{\bibinfo{volume}{113}}, \bibinfo{pages}{166402}
  (\bibinfo{year}{2014}).

\bibitem{HFA11}
\bibinfo{author}{Hohenadler, M.}, \bibinfo{author}{Fehske, H.} \&
  \bibinfo{author}{Assaad, F.~F.}
\newblock \bibinfo{journal}{\bibinfo{title}{Dynamic charge correlations near
  the {P}eierls transition}}.
\newblock {\emph{\JournalTitle{Phys. Rev. B}}} \textbf{\bibinfo{volume}{83}},
  \bibinfo{pages}{115105} (\bibinfo{year}{2011}).

\bibitem{WAH16}
\bibinfo{author}{Weber, M.}, \bibinfo{author}{Assaad, F.~F.} \&
  \bibinfo{author}{Hohenadler, M.}
\newblock \bibinfo{journal}{\bibinfo{title}{Continuous-time quantum {M}onte
  {C}arlo for fermion-boson lattice models: Improved bosonic estimators and
  application to the {H}olstein model}}.
\newblock {\emph{\JournalTitle{Phys. Rev. B}}} \textbf{\bibinfo{volume}{94}},
  \bibinfo{pages}{245138} (\bibinfo{year}{2016}).

\bibitem{JF07}
\bibinfo{author}{Jeckelmann, E.} \& \bibinfo{author}{Fehske, H.}
\newblock \bibinfo{journal}{\bibinfo{title}{Exact numerical methods for
  electron-phonon problems}}.
\newblock {\emph{\JournalTitle{Rivista del Nuovo Cimento}}}
  \textbf{\bibinfo{volume}{30}}, \bibinfo{pages}{259} (\bibinfo{year}{2007}).

\bibitem{To50}
\bibinfo{author}{Tomonaga, S.}
\newblock \bibinfo{journal}{\bibinfo{title}{Remarks on {B}loch's method of
  sound waves applied to many-fermion problems}}.
\newblock {\emph{\JournalTitle{Prog. Theor. Phys.}}}
  \textbf{\bibinfo{volume}{5}}, \bibinfo{pages}{544} (\bibinfo{year}{1950}).

\bibitem{Lu63}
\bibinfo{author}{Luttinger, J.~M.}
\newblock \bibinfo{journal}{\bibinfo{title}{An exactly soluble model of a
  many‐fermion system}}.
\newblock {\emph{\JournalTitle{J. Math. Phys.}}} \textbf{\bibinfo{volume}{4}},
  \bibinfo{pages}{1154} (\bibinfo{year}{1963}).

\bibitem{ZFA89}
\bibinfo{author}{Zheng, H.}, \bibinfo{author}{Feinberg, D.} \&
  \bibinfo{author}{Avignon, M.}
\newblock \bibinfo{journal}{\bibinfo{title}{Effect of quantum fluctuations on
  the {Peierls} dimerization in the one-dimensional molecular-crystal model}}.
\newblock {\emph{\JournalTitle{Phys. Rev. B}}} \textbf{\bibinfo{volume}{39}},
  \bibinfo{pages}{9405} (\bibinfo{year}{1989}).

\bibitem{WF98b}
\bibinfo{author}{Wei{\ss}e, A.} \& \bibinfo{author}{Fehske, H.}
\newblock \bibinfo{journal}{\bibinfo{title}{{Peierls} instability and optical
  response in the one-dimensional half-filled {Holstein} model of spinless
  fermions}}.
\newblock {\emph{\JournalTitle{Phys. Rev. B}}} \textbf{\bibinfo{volume}{58}},
  \bibinfo{pages}{13526} (\bibinfo{year}{1998}).

\bibitem{MHM96}
\bibinfo{author}{McKenzie, R.~H.}, \bibinfo{author}{Hamer, C.~J.} \&
  \bibinfo{author}{Murray, D.~W.}
\newblock \bibinfo{journal}{\bibinfo{title}{Quantum {M}onte {C}arlo study of
  the one-dimensional {H}olstein model of spinless fermions}}.
\newblock {\emph{\JournalTitle{Phys. Rev. B}}} \textbf{\bibinfo{volume}{53}},
  \bibinfo{pages}{9676} (\bibinfo{year}{1996}).

\bibitem{BMH98}
\bibinfo{author}{Bursill, R.~J.}, \bibinfo{author}{McKenzie, R.~H.} \&
  \bibinfo{author}{Hamer, C.~J.}
\newblock \bibinfo{journal}{\bibinfo{title}{Phase diagram of the
  one-dimensional {H}olstein model of spinless fermions}}.
\newblock {\emph{\JournalTitle{Phys. Rev. Lett.}}}
  \textbf{\bibinfo{volume}{80}}, \bibinfo{pages}{5607} (\bibinfo{year}{1998}).

\bibitem{HWBAF06}
\bibinfo{author}{Hohenadler, M.}, \bibinfo{author}{Wellein, G.},
  \bibinfo{author}{Bishop, A.~R.}, \bibinfo{author}{Alvermann, A.} \&
  \bibinfo{author}{Fehske, H.}
\newblock \bibinfo{journal}{\bibinfo{title}{Spectral signatures of the
  {L}uttinger liquid to the charge-density-wave transition}}.
\newblock {\emph{\JournalTitle{Phys. Rev. B}}} \textbf{\bibinfo{volume}{73}},
  \bibinfo{pages}{245120} (\bibinfo{year}{2006}).

\bibitem{EF09a}
\bibinfo{author}{Ejima, S.} \& \bibinfo{author}{Fehske, H.}
\newblock \bibinfo{journal}{\bibinfo{title}{Luttinger parameters and momentum
  distribution function for the half-filled spinless fermion {H}olstein model:
  A {DMRG} approach}}.
\newblock {\emph{\JournalTitle{Europhys. Lett.}}}
  \textbf{\bibinfo{volume}{87}}, \bibinfo{pages}{27001} (\bibinfo{year}{2009}).

\bibitem{AK99}
\bibinfo{author}{Alexandrov, A.~S.} \& \bibinfo{author}{Kornilovitch, P.~E.}
\newblock \bibinfo{journal}{\bibinfo{title}{Mobile small polaron}}.
\newblock {\emph{\JournalTitle{Phys. Rev. Lett.}}}
  \textbf{\bibinfo{volume}{82}}, \bibinfo{pages}{807} (\bibinfo{year}{1999}).

\bibitem{FLW00}
\bibinfo{author}{Fehske, H.}, \bibinfo{author}{Loos, J.} \&
  \bibinfo{author}{Wellein, G.}
\newblock \bibinfo{journal}{\bibinfo{title}{Lattice polaron formation: Effects
  of non-screened electron-phonon interaction}}.
\newblock {\emph{\JournalTitle{Phys. Rev. B}}} \textbf{\bibinfo{volume}{61}},
  \bibinfo{pages}{8016} (\bibinfo{year}{2000}).

\bibitem{Ho16}
\bibinfo{author}{Hohenadler, M.}
\newblock \bibinfo{journal}{\bibinfo{title}{Interplay of site and bond
  electron-phonon coupling in one dimension}}.
\newblock {\emph{\JournalTitle{Phys. Rev. Lett.}}}
  \textbf{\bibinfo{volume}{117}}, \bibinfo{pages}{206404}
  (\bibinfo{year}{2016}).

\bibitem{MB13}
\bibinfo{author}{Marchand, D. J.~J.} \& \bibinfo{author}{Berciu, M.}
\newblock \bibinfo{journal}{\bibinfo{title}{Effect of dispersive optical
  phonons on the behavior of a {H}olstein polaron}}.
\newblock {\emph{\JournalTitle{Phys. Rev. B}}} \textbf{\bibinfo{volume}{88}},
  \bibinfo{pages}{060301} (\bibinfo{year}{2013}).

\bibitem{BT21}
\bibinfo{author}{Bon\ifmmode~\check{c}\else \v{c}\fi{}a, J.} \&
  \bibinfo{author}{Trugman, S.~A.}
\newblock \bibinfo{journal}{\bibinfo{title}{Dynamic properties of a polaron
  coupled to dispersive optical phonons}}.
\newblock {\emph{\JournalTitle{Phys. Rev. B}}} \textbf{\bibinfo{volume}{103}},
  \bibinfo{pages}{054304} (\bibinfo{year}{2021}).

\bibitem{BT22}
\bibinfo{author}{Bon\ifmmode~\check{c}\else \v{c}\fi{}a, J.} \&
  \bibinfo{author}{Trugman, S.~A.}
\newblock \bibinfo{journal}{\bibinfo{title}{Electron removal spectral function
  of a polaron coupled to dispersive optical phonons}}.
\newblock {\emph{\JournalTitle{Phys. Rev. B}}} \textbf{\bibinfo{volume}{106}},
  \bibinfo{pages}{174303} (\bibinfo{year}{2022}).

\bibitem{JBH22}
\bibinfo{author}{Jansen, D.}, \bibinfo{author}{Bon\ifmmode~\check{c}\else
  \v{c}\fi{}a, J.} \& \bibinfo{author}{Heidrich-Meisner, F.}
\newblock \bibinfo{journal}{\bibinfo{title}{Finite-temperature optical
  conductivity with density-matrix renormalization group methods for the
  {H}olstein polaron and bipolaron with dispersive phonons}}.
\newblock {\emph{\JournalTitle{Phys. Rev. B}}} \textbf{\bibinfo{volume}{106}},
  \bibinfo{pages}{155129} (\bibinfo{year}{2022}).

\bibitem{CF24}
\bibinfo{author}{Chakraborty, M.} \& \bibinfo{author}{Fehske, H.}
\newblock \bibinfo{journal}{\bibinfo{title}{Quantum transport in an environment
  parametrized by dispersive bosons}}.
\newblock {\emph{\JournalTitle{Phys. Rev. B}}} \textbf{\bibinfo{volume}{109}},
  \bibinfo{pages}{085125} (\bibinfo{year}{2024}).

\bibitem{KB24}
\bibinfo{author}{Kova\ifmmode~\check{c}\else \v{c}\fi{}, K.} \&
  \bibinfo{author}{Bon\ifmmode~\check{c}\else \v{c}\fi{}a, J.}
\newblock \bibinfo{journal}{\bibinfo{title}{Light bipolarons in a system of
  electrons coupled to dispersive optical phonons}}.
\newblock {\emph{\JournalTitle{Phys. Rev. B}}} \textbf{\bibinfo{volume}{109}},
  \bibinfo{pages}{064304} (\bibinfo{year}{2024}).

\bibitem{Ed06}
\bibinfo{author}{Edwards, D.~M.}
\newblock \bibinfo{journal}{\bibinfo{title}{A quantum phase transition in a
  model with boson-controlled hopping}}.
\newblock {\emph{\JournalTitle{Physica B}}} \textbf{\bibinfo{volume}{378-380}},
  \bibinfo{pages}{133} (\bibinfo{year}{2006}).

\bibitem{AEF07}
\bibinfo{author}{Alvermann, A.}, \bibinfo{author}{Edwards, D.~M.} \&
  \bibinfo{author}{Fehske, H.}
\newblock \bibinfo{journal}{\bibinfo{title}{Boson-controlled quantum
  transport}}.
\newblock {\emph{\JournalTitle{Phys. Rev. Lett.}}}
  \textbf{\bibinfo{volume}{98}}, \bibinfo{pages}{056602}
  (\bibinfo{year}{2007}).

\bibitem{WFAE08}
\bibinfo{author}{Wellein, G.}, \bibinfo{author}{Fehske, H.},
  \bibinfo{author}{Alvermann, A.} \& \bibinfo{author}{Edwards, D.~M.}
\newblock \bibinfo{journal}{\bibinfo{title}{Correlation-induced metal insulator
  transition in a two-channel fermion-boson model}}.
\newblock {\emph{\JournalTitle{Phys. Rev. Lett.}}}
  \textbf{\bibinfo{volume}{101}}, \bibinfo{pages}{136402}
  (\bibinfo{year}{2008}).

\bibitem{EHF09}
\bibinfo{author}{Ejima, S.}, \bibinfo{author}{Hager, G.} \&
  \bibinfo{author}{Fehske, H.}
\newblock \bibinfo{journal}{\bibinfo{title}{Quantum phase transition in a {1D}
  transport model with boson affected hopping: {L}uttinger liquid versus
  charge-density-wave behavior}}.
\newblock {\emph{\JournalTitle{Phys. Rev. Lett.}}}
  \textbf{\bibinfo{volume}{102}}, \bibinfo{pages}{106404}
  (\bibinfo{year}{2009}).

\bibitem{LWF24}
\bibinfo{author}{Lange, F.}, \bibinfo{author}{Wellein, G.} \&
  \bibinfo{author}{Fehske, H.}
\newblock \bibinfo{journal}{\bibinfo{title}{Charge-order melting in the
  one-dimensional {E}dwards model}}.
\newblock {\emph{\JournalTitle{Phys. Rev. Res.}}} \textbf{\bibinfo{volume}{6}},
  \bibinfo{pages}{L022007} (\bibinfo{year}{2024}).

\bibitem{Sch11}
\bibinfo{author}{Schollw\"{o}ck, U.}
\newblock \bibinfo{journal}{\bibinfo{title}{The density-matrix renormalization
  group in the age of matrix product states}}.
\newblock {\emph{\JournalTitle{Ann. Phys.}}} \textbf{\bibinfo{volume}{326}},
  \bibinfo{pages}{96--192} (\bibinfo{year}{2011}).

\bibitem{Wh92}
\bibinfo{author}{White, S.~R.}
\newblock \bibinfo{journal}{\bibinfo{title}{Density matrix formulation for
  quantum renormalization groups}}.
\newblock {\emph{\JournalTitle{Phys. Rev. Lett.}}}
  \textbf{\bibinfo{volume}{69}}, \bibinfo{pages}{2863} (\bibinfo{year}{1992}).

\bibitem{ZVFVH18}
\bibinfo{author}{Zauner-Stauber, V.}, \bibinfo{author}{Vanderstraeten, L.},
  \bibinfo{author}{Fishman, M.~T.}, \bibinfo{author}{Verstraete, F.} \&
  \bibinfo{author}{Haegeman, J.}
\newblock \bibinfo{journal}{\bibinfo{title}{Variational optimization algorithms
  for uniform matrix product states}}.
\newblock {\emph{\JournalTitle{Phys. Rev. B}}} \textbf{\bibinfo{volume}{97}},
  \bibinfo{pages}{045145} (\bibinfo{year}{2018}).

\bibitem{HLOVV16}
\bibinfo{author}{Haegeman, J.}, \bibinfo{author}{Lubich, C.},
  \bibinfo{author}{Oseledets, I.}, \bibinfo{author}{Vandereycken, B.} \&
  \bibinfo{author}{Verstraete, F.}
\newblock \bibinfo{journal}{\bibinfo{title}{Unifying time evolution and
  optimization with matrix product states}}.
\newblock {\emph{\JournalTitle{Phys. Rev. B}}} \textbf{\bibinfo{volume}{94}},
  \bibinfo{pages}{165116} (\bibinfo{year}{2016}).

\bibitem{GLvD23}
\bibinfo{author}{Gleis, A.}, \bibinfo{author}{Li, J.-W.} \&
  \bibinfo{author}{von Delft, J.}
\newblock \bibinfo{journal}{\bibinfo{title}{Controlled bond expansion for
  density matrix renormalization group ground state search at single-site
  costs}}.
\newblock {\emph{\JournalTitle{Phys. Rev. Lett.}}}
  \textbf{\bibinfo{volume}{130}}, \bibinfo{pages}{246402}
  (\bibinfo{year}{2023}).

\bibitem{LGvD22}
\bibinfo{author}{Li, J.-W.}, \bibinfo{author}{Gleis, A.} \&
  \bibinfo{author}{von Delft, J.}
\newblock \bibinfo{title}{Time-dependent variational principle with controlled
  bond expansion for matrix product states} (\bibinfo{year}{2022}).
\newblock \bibinfo{note}{{arXiv:2208.10972}}.

\bibitem{PVM12}
\bibinfo{author}{Phien, H.~N.}, \bibinfo{author}{Vidal, G.} \&
  \bibinfo{author}{McCulloch, I.~P.}
\newblock \bibinfo{journal}{\bibinfo{title}{Infinite boundary conditions for
  matrix product state calculations}}.
\newblock {\emph{\JournalTitle{Phys. Rev. B}}} \textbf{\bibinfo{volume}{86}},
  \bibinfo{pages}{245107} (\bibinfo{year}{2012}).

\bibitem{MHOV13}
\bibinfo{author}{Milsted, A.}, \bibinfo{author}{Haegeman, J.},
  \bibinfo{author}{Osborne, T.~J.} \& \bibinfo{author}{Verstraete, F.}
\newblock \bibinfo{journal}{\bibinfo{title}{Variational matrix product ansatz
  for nonuniform dynamics in the thermodynamic limit}}.
\newblock {\emph{\JournalTitle{Phys. Rev. B}}} \textbf{\bibinfo{volume}{88}},
  \bibinfo{pages}{155116} (\bibinfo{year}{2013}).

\bibitem{ZGEN15}
\bibinfo{author}{Zauner, V.}, \bibinfo{author}{Ganahl, M.},
  \bibinfo{author}{Evertz, H.~G.} \& \bibinfo{author}{Nishino, T.}
\newblock \bibinfo{journal}{\bibinfo{title}{Time evolution within a comoving
  window: scaling of signal fronts and magnetization plateaus after a local
  quench in quantum spin chains}}.
\newblock {\emph{\JournalTitle{J. Phys.: Condens. Matter}}}
  \textbf{\bibinfo{volume}{27}}, \bibinfo{pages}{425602}
  (\bibinfo{year}{2015}).

\bibitem{SKMJHP21}
\bibinfo{author}{Stolpp, J.} \emph{et~al.}
\newblock \bibinfo{journal}{\bibinfo{title}{{Comparative study of
  state-of-the-art matrix-product-state methods for lattice models with large
  local Hilbert spaces without U(1) symmetry}}}.
\newblock {\emph{\JournalTitle{Computer Physics Communications}}}
  \textbf{\bibinfo{volume}{269}}, \bibinfo{pages}{108106}
  (\bibinfo{year}{2021}).

\bibitem{JW98b}
\bibinfo{author}{Jeckelmann, E.} \& \bibinfo{author}{White, S.~R.}
\newblock \bibinfo{journal}{\bibinfo{title}{Density-matrix
  renormalization-group study of the polaron problem in the {H}olstein model}}.
\newblock {\emph{\JournalTitle{Phys. Rev. B}}} \textbf{\bibinfo{volume}{57}},
  \bibinfo{pages}{6376} (\bibinfo{year}{1998}).

\bibitem{LF62}
\bibinfo{author}{Lang, I.~G.} \& \bibinfo{author}{Firsov, Y.~A.}
\newblock \bibinfo{journal}{\bibinfo{title}{Kinetic theory of semiconductors
  with low mobility}}.
\newblock {\emph{\JournalTitle{Zh. Eksp. Teor. Fiz.}}}
  \textbf{\bibinfo{volume}{43}}, \bibinfo{pages}{1843} (\bibinfo{year}{1962}).

\bibitem{DDY05}
\bibinfo{author}{Datta, S.}, \bibinfo{author}{Das, A.} \&
  \bibinfo{author}{Yarlagadda, S.}
\newblock \bibinfo{journal}{\bibinfo{title}{Many-polaron effects in the
  {H}olstein model}}.
\newblock {\emph{\JournalTitle{PRB}}} \textbf{\bibinfo{volume}{71}},
  \bibinfo{pages}{235118} (\bibinfo{year}{2005}).

\bibitem{Gi03}
\bibinfo{author}{Giamarchi, T.}
\newblock \emph{\bibinfo{title}{Quantum Physics in One Dimension}}
  (\bibinfo{publisher}{Clerendon Press}, \bibinfo{address}{Oxford},
  \bibinfo{year}{2003}).

\bibitem{KFM92}
\bibinfo{author}{Kane, C.~L.} \& \bibinfo{author}{Fisher, M. P.~A.}
\newblock \bibinfo{journal}{\bibinfo{title}{Transmission through barriers and
  resonant tunneling in an interacting one-dimensional electron gas}}.
\newblock {\emph{\JournalTitle{Phys. Rev. B}}} \textbf{\bibinfo{volume}{46}},
  \bibinfo{pages}{15233} (\bibinfo{year}{1992}).

\bibitem{KLOKC21}
\bibinfo{author}{Kang, Y.-T.}, \bibinfo{author}{Lo, C.-Y.},
  \bibinfo{author}{Oshikawa, M.}, \bibinfo{author}{Kao, Y.-J.} \&
  \bibinfo{author}{Chen, P.}
\newblock \bibinfo{journal}{\bibinfo{title}{Two-wire junction of inequivalent
  {T}omonaga-{L}uttinger liquids}}.
\newblock {\emph{\JournalTitle{Phys. Rev. B}}} \textbf{\bibinfo{volume}{104}},
  \bibinfo{pages}{235142} (\bibinfo{year}{2021}).

\bibitem{GS89}
\bibinfo{author}{Giamarchi, T.} \& \bibinfo{author}{Schulz, H.~J.}
\newblock \bibinfo{journal}{\bibinfo{title}{Correlation functions of
  one-dimensional quantum systems}}.
\newblock {\emph{\JournalTitle{Phys. Rev. B}}} \textbf{\bibinfo{volume}{39}},
  \bibinfo{pages}{4620} (\bibinfo{year}{1989}).

\bibitem{EGN05}
\bibinfo{author}{Ejima, S.}, \bibinfo{author}{Gebhard, F.} \&
  \bibinfo{author}{Nishimoto, S.}
\newblock \bibinfo{journal}{\bibinfo{title}{{T}omonaga-{L}uttinger parameters
  for doped {M}ott insulators}}.
\newblock {\emph{\JournalTitle{Europhys. Lett.}}}
  \textbf{\bibinfo{volume}{70}}, \bibinfo{pages}{492} (\bibinfo{year}{2005}).

\bibitem{KM12}
\bibinfo{author}{Karrasch, C.} \& \bibinfo{author}{Moore, J.~E.}
\newblock \bibinfo{journal}{\bibinfo{title}{Luttinger liquid physics from the
  infinite-system density matrix renormalization group}}.
\newblock {\emph{\JournalTitle{Phys. Rev. B}}} \textbf{\bibinfo{volume}{86}},
  \bibinfo{pages}{155156} (\bibinfo{year}{2012}).

\bibitem{HAF12}
\bibinfo{author}{Hohenadler, M.}, \bibinfo{author}{Assaad, F.~F.} \&
  \bibinfo{author}{Fehske, H.}
\newblock \bibinfo{journal}{\bibinfo{title}{Effect of electron-phonon
  interaction range for a half-filled band in one dimension}}.
\newblock {\emph{\JournalTitle{Phys. Rev. Lett.}}}
  \textbf{\bibinfo{volume}{109}}, \bibinfo{pages}{116407}
  (\bibinfo{year}{2012}).

\bibitem{OLSA91}
\bibinfo{author}{Ogata, M.}, \bibinfo{author}{Luchini, M.~U.},
  \bibinfo{author}{Sorella, S.} \& \bibinfo{author}{Assaad, F.~F.}
\newblock \bibinfo{journal}{\bibinfo{title}{Phase diagram of the
  one-dimensional $t$-$j$ model}}.
\newblock {\emph{\JournalTitle{Phys. Rev. Lett.}}}
  \textbf{\bibinfo{volume}{66}}, \bibinfo{pages}{2388} (\bibinfo{year}{1991}).

\bibitem{ESBF12}
\bibinfo{author}{Ejima, S.}, \bibinfo{author}{Sykora, S.},
  \bibinfo{author}{Becker, K.~W.} \& \bibinfo{author}{Fehske, H.}
\newblock \bibinfo{journal}{\bibinfo{title}{Phase separation in the {E}dwards
  model}}.
\newblock {\emph{\JournalTitle{Phys. Rev. B}}} \textbf{\bibinfo{volume}{86}},
  \bibinfo{pages}{155149} (\bibinfo{year}{2012}).

\bibitem{GMSR21b}
\bibinfo{author}{Gotta, L.}, \bibinfo{author}{Mazza, L.},
  \bibinfo{author}{Simon, P.} \& \bibinfo{author}{Roux, G.}
\newblock \bibinfo{journal}{\bibinfo{title}{Two-fluid coexistence in a spinless
  fermions chain with pair hopping}}.
\newblock {\emph{\JournalTitle{Phys. Rev. Lett.}}}
  \textbf{\bibinfo{volume}{126}}, \bibinfo{pages}{206805}
  (\bibinfo{year}{2021}).

\bibitem{itensor}
\bibinfo{author}{Fishman, M.}, \bibinfo{author}{White, S.~R.} \&
  \bibinfo{author}{Stoudenmire, E.~M.}
\newblock \bibinfo{journal}{\bibinfo{title}{{The ITensor Software Library for
  Tensor Network Calculations}}}.
\newblock {\emph{\JournalTitle{SciPost Phys. Codebases}}} \bibinfo{pages}{4}
  (\bibinfo{year}{2022}).

\bibitem{itensor-r0.3}
\bibinfo{author}{Fishman, M.}, \bibinfo{author}{White, S.~R.} \&
  \bibinfo{author}{Stoudenmire, E.~M.}
\newblock \bibinfo{journal}{\bibinfo{title}{{Codebase release 0.3 for
  ITensor}}}.
\newblock {\emph{\JournalTitle{SciPost Phys. Codebases}}}
  \bibinfo{pages}{4--r0.3} (\bibinfo{year}{2022}).

\end{thebibliography}

%For data citations of datasets uploaded to e.g. \emph{figshare}, please use the \verb|howpublished| option in the bib entry to specify the platform and the link, as in the \verb|Hao:gidmaps:2014| example in the sample bibliography file.

\section*{Acknowledgements}
The authors acknowledge the scientific support
and HPC resources provided by the Erlangen National High
Performance Computing Center (NHR@FAU) of the Friedrich-Alexander-Universität Erlangen-Nürnberg (FAU). The hardware is funded by the German Research Foundation (DFG). MPS simulations were performed using the ITensor library\cite{itensor,itensor-r0.3}.

\section*{Author contributions statement}
F.L. and H.F. contributed equally to this work.

\section*{Additional information}

\textbf{Competing interests} The authors declare no competing financial interests.

%The corresponding author is responsible for submitting a \href{http://www.nature.com/srep/policies/index.html#competing}{competing interests statement} on behalf of all authors of the paper. This statement must be included in the submitted article file.

\end{document}